\documentclass{aa}
\usepackage{natbib,amssymb,graphicx,graphics}
\newcommand{\excs}{\extracolsep{\fill}}
\begin{document}
\title{Planetary systems in close binary stars: the case of HD\,196885}
\subtitle{Combined astrometric and radial velocity study}
\titlerunning{Dynamical Study of HD\,196885\,AB}

\author{G. Chauvin\inst{1}
	\and H. Beust\inst{1}
	\and A.-M. Lagrange\inst{1}
	\and A. Eggenberger \inst{1}
}
\institute{
$^{1}$Laboratoire d'Astrophysique, Observatoire de Grenoble, UJF, CNRS; BP 53,
  F-38041 GRENOBLE Cedex 9 (France)\\
}

\offprints{G. Chauvin}
\date{Received ; accepted }

  \abstract 
  {More than fifty \textbf{candidate} planets are presently known to orbit one component
    of a binary or a multiple star system. Planets can therefore form
    and survive in such an environment, although recent observing surveys
    indicate that short-separation binaries do not favour the presence
    of a planetary system around one of the component. Dynamical
    interactions with the secondary component can actually
    significantly impact the giant planet formation and evolution. For
    this reason, rare close binaries hosting giant planets offer
    an ideal laboratory to explore the properties and the stability of
    such extreme planetary systems.}
  {In the course of our CFHT and VLT coronographic imaging survey
  dedicated to the search for faint companions of exoplanet host
  stars, a close ($\sim20$~AU) secondary stellar companion to the exoplanet
  host HD196885\,A was discovered. In this study, our aim is to monitor
  the orbital motion of the binary companion. Combining radial velocity
  and high contrast imaging observations, we aim to derive the orbital
  properties of the complete system and to test its dynamical
  stability to reveal its formation.}
  {For more than 4 years, we have used the NaCo near-infrared adaptive
    optics instrument to monitor the astrometric position of
    HD\,196885\,B relative to A. The system was observed at five
    different epochs from August 2005 to August 2009 and accurate
    relative positions were determined.}
  {Our observations fully reject the \textbf{stationary} background
    hypothesis for HD196885\,B. The two components are found to be
    comoving. The orbital motion of HD196885\,B is well resolved and
    the orbital curvature is even detected. From our imaging data
    combined with published radial velocity measurements, we refine
    the complete orbital \textbf{parameters} of the stellar component.
    We derive for the first time its orbital inclination and its
    accurate mass. We find also solutions for the inner giant planet
    HD196885 Ab compatible with previous independent radial velocity
    studies. Finally, we investigate the stability of the inner giant
    planet HD196885\,Ab due to the binary companion proximity. Our
    dynamical simulations show that the system is currently and
    surprisingly more stable in a high mutual inclination
    configuration that falls in the Kozai resonance regime.  If
    confirmed, this system would constitute one of the most compact
    non-coplanar systems known so far. It would raise several
    questions about its formation and stability.}
{}

   \maketitle

\section{Introduction}


Among current exoplanets hunting techniques, radial velocity (RV)
measurements are nowadays the most successful method for detecting
exo-planetary systems (Udry \& Santos 2007; Cumming et
al. 2008). \textbf{Although the minimum mass is derived, a small
  correction including the distribution of inclination is expected to
  access the true mass distribution of the current candidate planets
  (Watson et al. 2010; Jorrissen et al. 2001)}. Originally focused on
quiet solar-type stars that show numerous thin absorption lines, RV
surveys have recently diversified their samples to consider a broader
class of primary stars. Telluric planets are now preferentially
searched for around M dwarfs as the habitable zone can be explored
(Mayor et al. 2009; Charbonneau et al. 2009). Giant planets have
probably been discovered around intermediate mass objects (Lagrange et
al. 2009), giant stars (Doëllinger et al. 2009; Lovis et al. 2007) and
have actively been for earched around young active stars (Setiawan et
al. 2008). Finally, planets are now scrutinized in multiple stellar
systems where RV surveys used to exclude them. This was mainly
related to the difficulty of processing the multiple stellar RV
signals at the required precision to detect planets (Konacki 2005;
Eggenberger \& Udry 2007; Toyota et al. 2009; Desidera et
al. 2010). Binaries and triple are particularly interesting to test
the predictions of the planetary formation and evolution
processes. They enable us to understand how a perturber will impact the
planetary system formation and dynamical evolution. Originally,
planets were rapidly found in close spectroscopic binaries (Gliese~86,
Queloz et al. 2000; $\gamma$ Cep Hatzes et al. 2003), confirming that
circumstellar planetary systems could form and survive in such a
hostile environment (binary separation $\lesssim 20$~AU). Circumbinary
planets, undiscovered up until now, remain more difficult to detect due
to the limited size of the sample and only the recent start-up of dedicated
programmes.

A few years ago, studies and surveys had success in tackling the
problem of duplicity in planetary systems. Zucker \& Mazeh (2002)
initially showed that the planet properties in binaries were different
to planets around single stars. Eggenberger et al. (2004) confirmed
that the most massive short period planets were found in binary
systems and that planets with short ($P\le40$~days) periods in
binaries were likely to have low eccentricities. Tamuz et al. (2008)
finally found that extremely high eccentric planets were all in
binaries. In parallel to Doppler surveys, imaging surveys studied
the effect of duplicity on the planet occurence around longer period
($\ge100$~AU) binaries. Catalogs compiled by Raghavan et
al. (2006), Desidera \& Barbieri (2007) and Bonavita \& Desidera
(2007) did not find any significant discrepancies in the planet
frequency between binary and single stellar systems. Only high
contrast imaging studies have been able to access
intermediate-separation ($20-100$~AU) binaries where the companion
influence is the most expected. Several deep imaging studies have
revealed a number of additional close companions to known exoplanetary
hosts (Patience et al. 2002; Chauvin et al. 2006; Eggenberger et
al. 2007; Mugrauer et al. 2009). However, a dedicated survey using a
reference sample was mandatory to conduct a proper multiplicity study
to test the impact of duplicity on the giant planet
occurence. Eggenberger et al. (2007) observed two subsamples of more
than 50 stars each. They found a lower corrected binary fraction for
the planet-host subsample, particularly for physical separations
shorter than $100$~AU (Eggenberger et al. 2008). This statistical
result corroborates the theoretical prediction that multiplicity has a
negative impact on planetary formation and evolution at less than
$100$~AU (e.g. Nelson et al. 2000; Mayer et al. 2005; Th\'ebault et
al. 2006).

\textbf{Binaries with small semi-major axes ($\lesssim 20$~AU) hosting
  a planetary system offer an ideal opportunity} to characterize in
more detail the dynamical impact of the binary companion on the inner
circumstellar planet. They can also serve as a testbed for planet
formation models, as they push many of these models parameters
to their limits. We conducted such a study combining RV and AO imaging
observations for the Gliese~86 system (Lagrange et al. 2006). We
confirmed that the companion was a white dwarf identified in
spectrocopy by Neuh\"auser et al. (2005). We showed that the planetary
system around Gliese~86\,A had survived the later stages of evolutiion of the
white dwarf progenitor, a probable late-F to early-K type star (i.e
the mass loss of the B component and a semi-major axis intially
reduced to 13~AU compared with its 18~AU current value). In the course
of the VLT/CFHT deep imaging surveys of exoplanet hosts (Chauvin et
al. 2006, 2007), we discovered a close ($\sim 20$~AU) binary companion
to HD\,196885 (see Fig.~1). This system is well suited for a detailed
dynamical study. In this paper, we summarize the observations and the
results of our recent imaging campaign aimed at monitoring the binary
orbital motion.  We report the best orbital adjustment to fit the
combined imaging and RV observations of the complete system. Finally,
we discuss the dynamical stability of the inner planetary system that
has formed and survived despite the dynamical influence of the close
binary companion.

\begin{figure}[t]
\centering
\vspace{-0.1cm}
\includegraphics[width = \columnwidth]{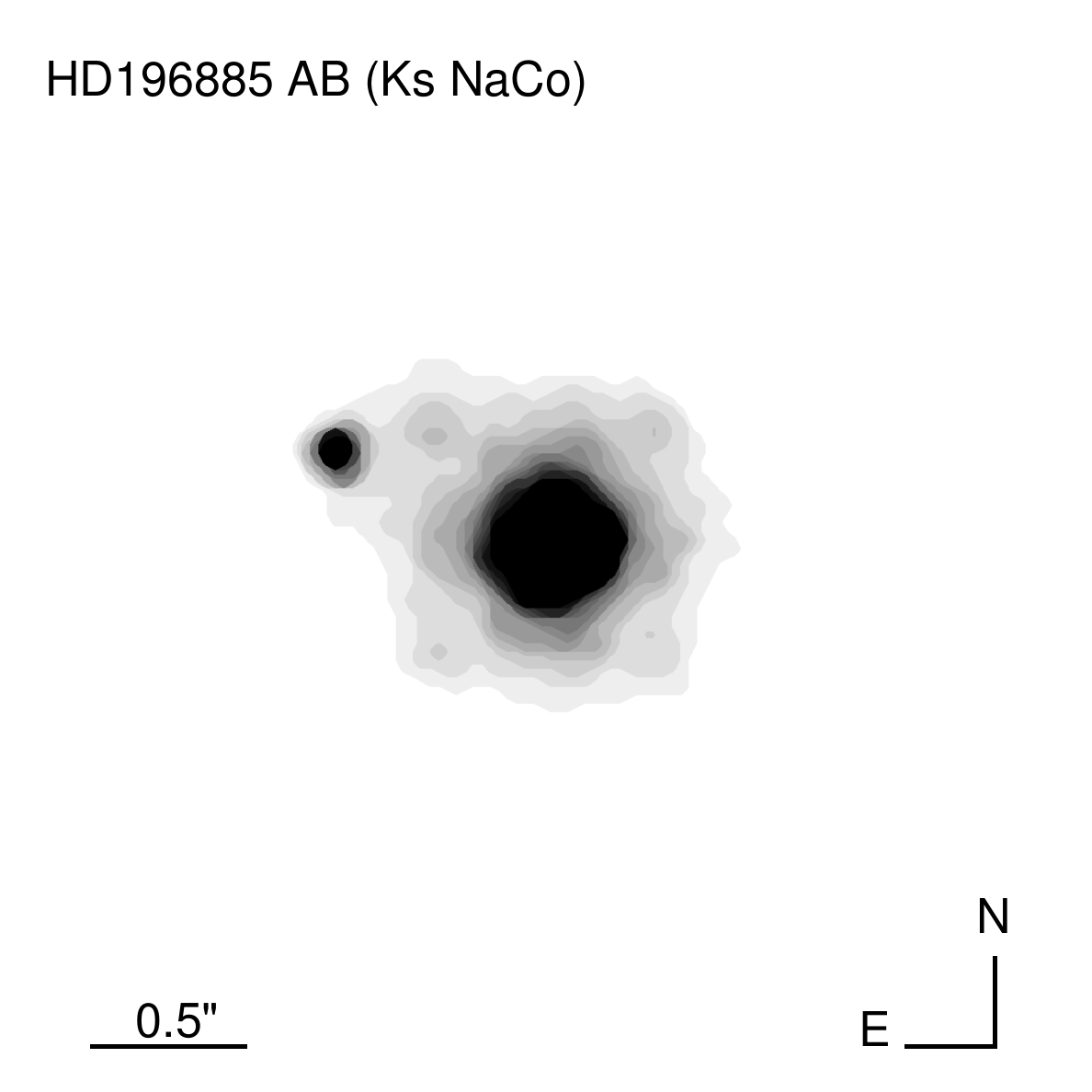}\hspace{0.5cm}
\caption{VLT/NACO image obtained in $K_s$-band using the S27 CONICA
  platescale. \textbf{The binary is clearly resolved}.}
\label{fig:image}
\end{figure}

\begin{table*}[t]
\caption{Relative astrometric position of the binary companion HD196885\,B to the exoplanet host star HD\,196885\,A.}
\label{tab:astro}
\centering
\begin{tabular*}{\textwidth}{@{\excs}lllllll}     
\hline\hline\noalign{\smallskip}
 UT Date     & Filter/Obj     & $\Delta\alpha$   & $\Delta\delta$    &    $\Delta K_s$      & Plate Scale    &   True North \\
             &            & (mas)            & (mas)             &    (mag)         & (mas)          &   (o)              \\
\noalign{\smallskip}\hline\noalign{\smallskip}
01/08/2005   & $K_s$/S27 & $658.5\pm1.7$ & $273.6\pm1.6$ & $3.08\pm0.05$ & $27.01\pm0.05$     & $0.09\pm0.12$    \\
26/08/2006   & $K_s$/S27 & $649.8\pm1.9$ & $292.6\pm2.4$ & $3.06\pm0.04$ & $27.01\pm0.05$     & $0.02\pm0.20$    \\
25/08/2007    & $K_s$/S27 & $640.2\pm3.1$ & $309.7\pm3.2$ & $3.07\pm0.06$ & $27.01\pm0.05$     & $0.05\pm0.15$    \\
28/06/2008    & $K_s$/S27 & $630.9\pm3.1$ & $326.3\pm3.2$ & $3.00\pm0.10$  & $27.01\pm0.05$     & $-0.25\pm0.14$    \\ 
27/08/2009    & $K_s$/S27 & $614.5\pm3.0$ & $342.1\pm2.9$ & $3.14\pm0.09$  & $27.01\pm0.05$     & $-0.35\pm0.08$    \\ 
\noalign{\smallskip}\hline
\end{tabular*}
\end{table*}

\section{The HD196885 exoplanet host binary}
\label{sec:description}


HD196885 is an F8V ($V=6.398$, $B-V=0.559$) star located at
$33.0\pm0.9$~pc \textbf{(Perryman et al. 1997)}. Based on CORALIE
spectra, Sousa et al. (2006) derived a spectrocopic temperature,
surface gravity and metallicity of $T_{\rm{eff}}=6340\pm39$~K,
log$(g)=4.46\pm0.02$ and [Fe/H]$ = 0.29\pm0.05$ respectively. The star
v$\,sin$i was estimated to $7.3\pm1.5$~kms$^{-1}$ from ELODIE spectra
(Correia et al. 2008). These results were recently supported by
independent Lick observations reported by Fischer et al. (2009). The
chromospheric activity level is relatively low.  Bolometric luminosity
correction and evolutionary model predictions lead to an estimate of
the luminosity and the mass of $2.4~L_{\odot}$ and $1.3~M_{\odot}$
respectively. Finally, the corresponding stellar age derived from
evolutionary tracks and from the activity level varies between 1.5 to
3.5~Gyr (see Correia et al. 2008; Fischer et al. 2009). As noted by
Correia et al. (2008), HD196885 might be part of a wider binary system
with the star BD+104351\,B\footnote{and not BD+104251\,B as mentionned
  by Correia et al. (2008)} located 192~$\!''$ north (that would
correspond to a minimum physical separation of 6330~AU, Dommanget et
al. 2002).

In 2004, a RV variation was measured and adjusted with a preliminary
orbital period of $P=0.95$~yrs. The result was temporarily reported on
the California Planet Search Exoplanet Web site and was thus withdrawn
due to a significant residual drift in the orbital
solution. Nevertheless, this star was included in our deep imaging
survey of stars hosting planets detected by RV observations (Chauvin
et al. 2006). Our imaging and spectroscopic observations led to the
detection of a co-moving M1$\pm$1V dwarf companion located at only
0.7~$\!''$ (23~AU in projected physical separation) and likely to be
responsible for the trend seen in the Lick RV residuals (Chauvin et
al. 2007). Our two epochs of observations resolved the binary orbital
motion confirming its physical nature.

Using a double-Keplerian model for the binary star and the planet to
adjust their ELODIE, CORALIE and CORAVEL observations spread over
14~years, Correia et al. (2008) derived a first range of orbital
solutions. They revised the planet solution with a minimum mass of
$M_{Ab}\,\rm{sin}i=2.96$~$M_{\rm{Jup}}$, a period of $P=3.69\pm0.03$~yrs
and an eccentricity of $e=0.462\pm0.026$. Moreover, they found
additional constraints for the binary companion HD196885\,B with a
period of $P> 40$~yr, a semi-major axis $a > 14$~AU and a minimum mass of
$M_B\,\rm{sin}i > 0.28~M_{\odot}$. Based on Lick observations, Fischer
et al. (2009) recently derived consistent results for both the inner
planet and the binary companion. Nevertheless, in both studies a
relatively large range of masses and periods remains for the
binary companion. Additional observing constraints are therefore
needed to refine the binary companion properties and to understand how
it could have affected the formation and the stability of the
circumstellar planetary system around HD\,196885\,A.

\begin{figure}[t]
\centering
\vspace{-0.1cm}
\includegraphics[width = \columnwidth]{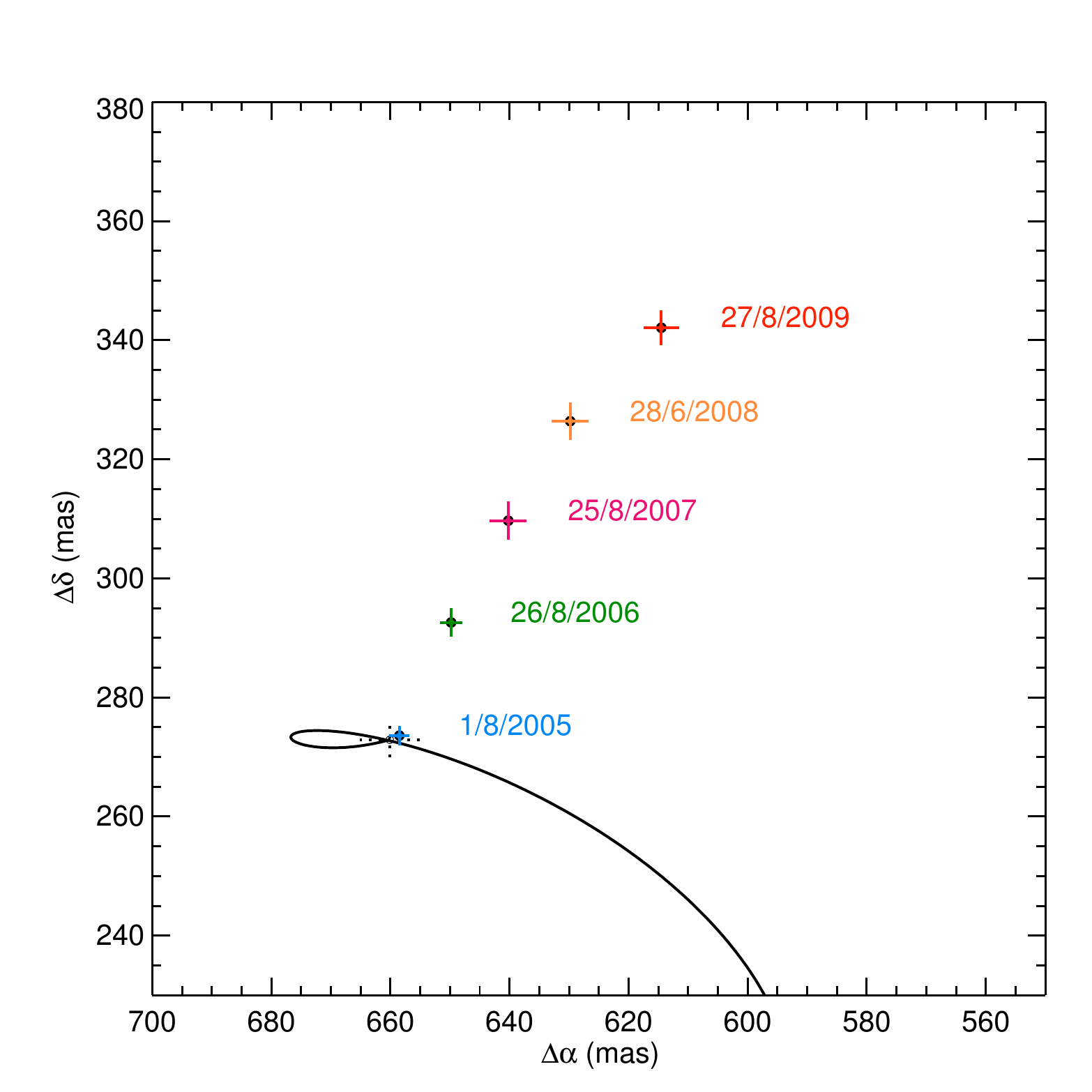}

\caption{VLT/NACO measurements with uncertainties in the offset
  positions of HD\,196885\,B relative to A, obtained in August 1st
  2005, August 28th 2006, August 27th 2007, June 26th 2008 and August
  27th 2009. \textbf{The \textit{solid line} gives the expected
    variation of the offset positions of B relative to A if B is a
    background stationary object. This variation takes into account
    the initial offset position of B relative to a in August 1st 2005
    and the proper and parallactic motions of HD\,196885\,A. After
    April 2006, the predicted offset positions go beyond the ($\Delta
    \alpha, \Delta \delta$) astrometric range considered for this
    figure.}}

\label{fig:astrodiag}
\end{figure}

\section{Observation and Data Reduction}
\label{sec:observations}

The orbit of HD\,196885\,AB was monitored with the NACO (NAOS-CONICA)
high contrast Adaptive Optics (AO) imager of the VLT-UT4. The NAOS AO
system (Rousset et al. 2002) is equipped with a tip-tilt mirror, a 185
piezo actuator deformable mirror and two wavefront sensors (Visible
and IR). Attached to NAOS, CONICA (Lenzen et al. 1998) is the near
infrared ($1-5\,\mu$m domain) imaging, Lyot coronagraphic,
spectroscopic and polarimetric camera, equipped with a
$1024\times1024$ pixels Aladdin InSb array. Observations were obtained
at five different epochs in August 1st 2005, August 26th 2006, August
25th 2007, June 28th 2008 and August 27th 2009. Over the different
observing campaigns, the atmospheric conditions were sufficiently
stable to close the AO loop and resolve both components. The offset
position of HD\,196885\,B relative to A was well monitored at each
epoch.  The typical observing sequence included a set of five jittered
images obtained using the K$_{s}$ filter and the S27 camera CONICA
(mean plate scale of 27.01~mas/pixel). Thsi led to a total exposure
time of $\sim5$~min on source. To calibrate the plate scale and the
detector orientation, we observed the astrometric field of
$\theta$\,Ori\,1\,C (McCaughrean \& Stauffer 1994).  When not
observable in June 2008, we used as secondary calibrator the
astrometric binary IDS\,21506S5133 (van Dessel \& Sinachopoulos 1993),
recalibrated with the $\theta$ Ori\,1 C field.

After cosmetic reductions (dead and hot pixels, dark and flat) using
\textit{eclipse} (Devillard 1997), we applied the deconvolution
algorithm of V\'eran \& Rigaut (1998) to obtain the offset position of
HD\,196885\,B relative to A at each epoch. Single stars of similar
brightness observed on the same night were used for point spread function
(PSF) estimation. The results are reported in
Table~\ref{tab:astro}. The platescale and the true North orientation
of the detector are given for each measurement. The relative positions
are shown in Fig.~\ref{fig:astrodiag}.  The expected variation of
offset positions, if B is a background stationary object, is shown
(\textit{solid line}). It takes into account the initial offset
position of B relative A and the proper
($(\mu_{\alpha},\mu_{\delta})=(47.5\pm0.9, 83.1\pm0.5)$~mas/yr) and
parallactic ($\pi=30.31\pm0.81$~mas) motions of HD\,196885\,A.  The
orbital curvature is detected and the background stationary hypothesis
for HD\,196885\,B is fully excluded.

\section{Orbital solution}

\begin{figure*}[t]
\centering
\includegraphics[width = 9.1cm]{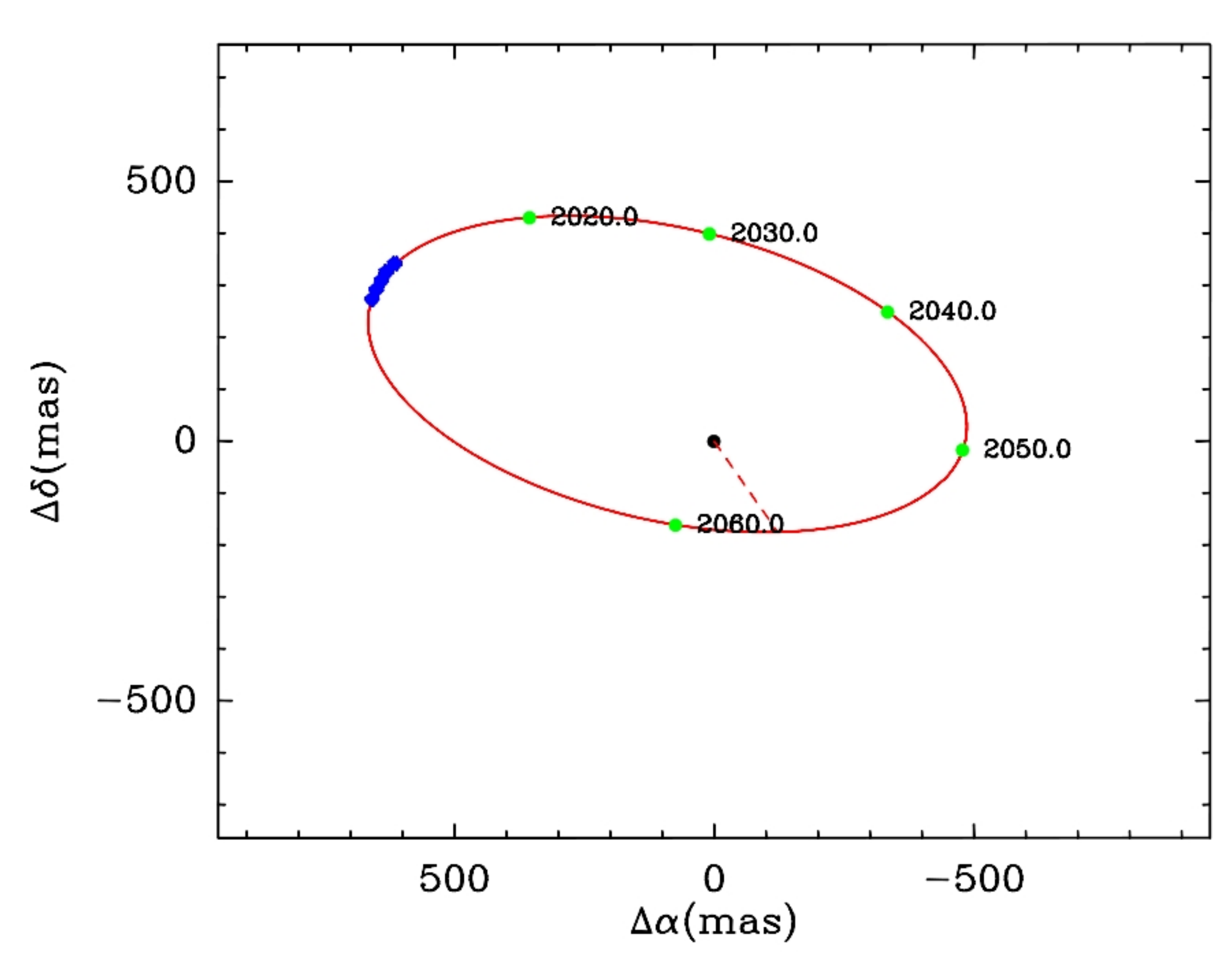}
\includegraphics[width = 9.1cm]{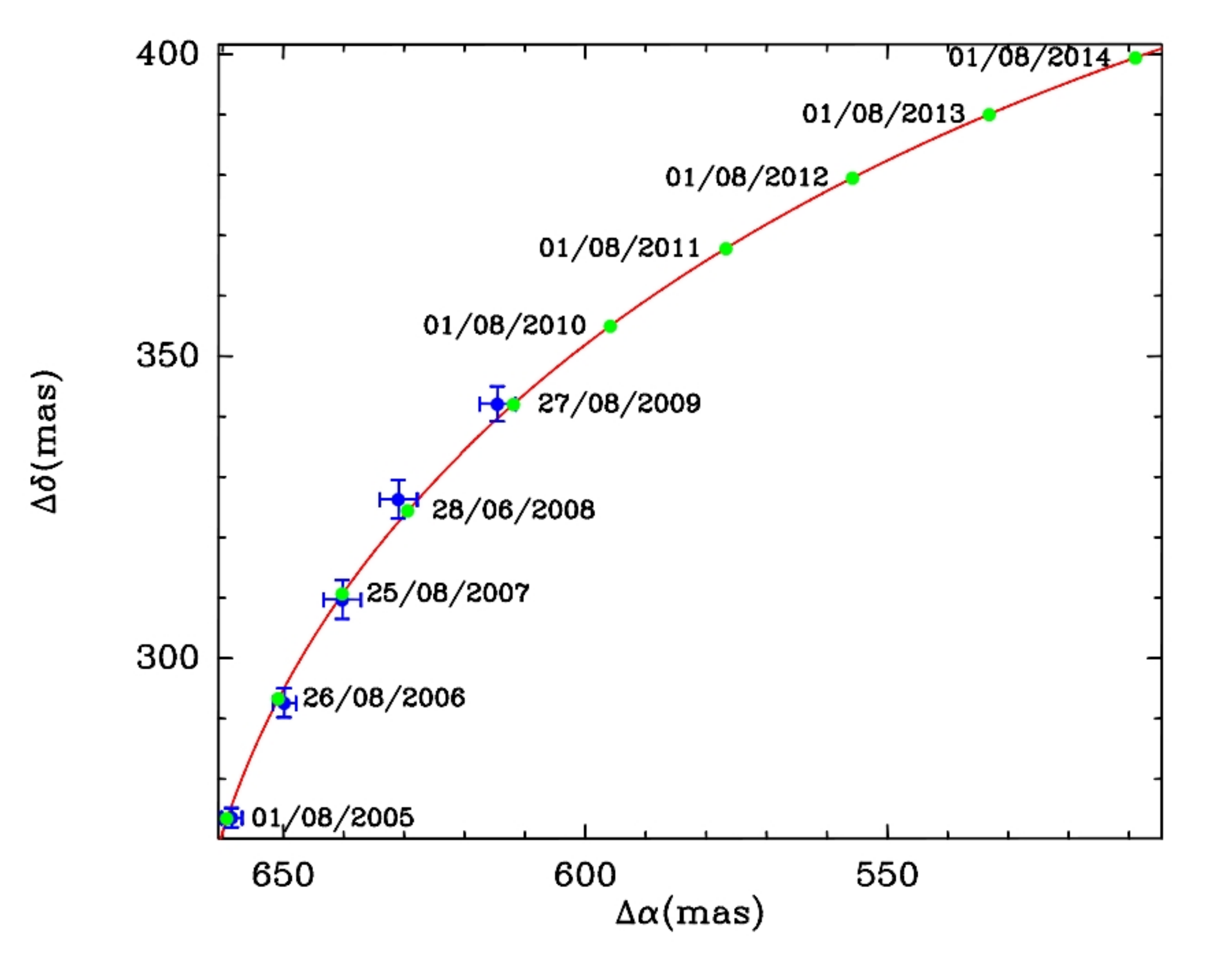}
\caption{NaCo astrometric observations and orbital solution of
  HD\,196885\,A and B.  The full orbital solution is sketched in
  \textit{red} and superimposed to the astrometric data points in
  \textit{blue}. Predicted positions from the fit are shown in
  \textit{green}. \textbf{Left,} Full orbital
  solution. \textbf{Right,} Zoomed on with the astometric data points
  and their uncertainties. }
\label{fig:orbitfit1}
\end{figure*}
\begin{figure*}[t]
\centering
\includegraphics[width = 8.9cm]{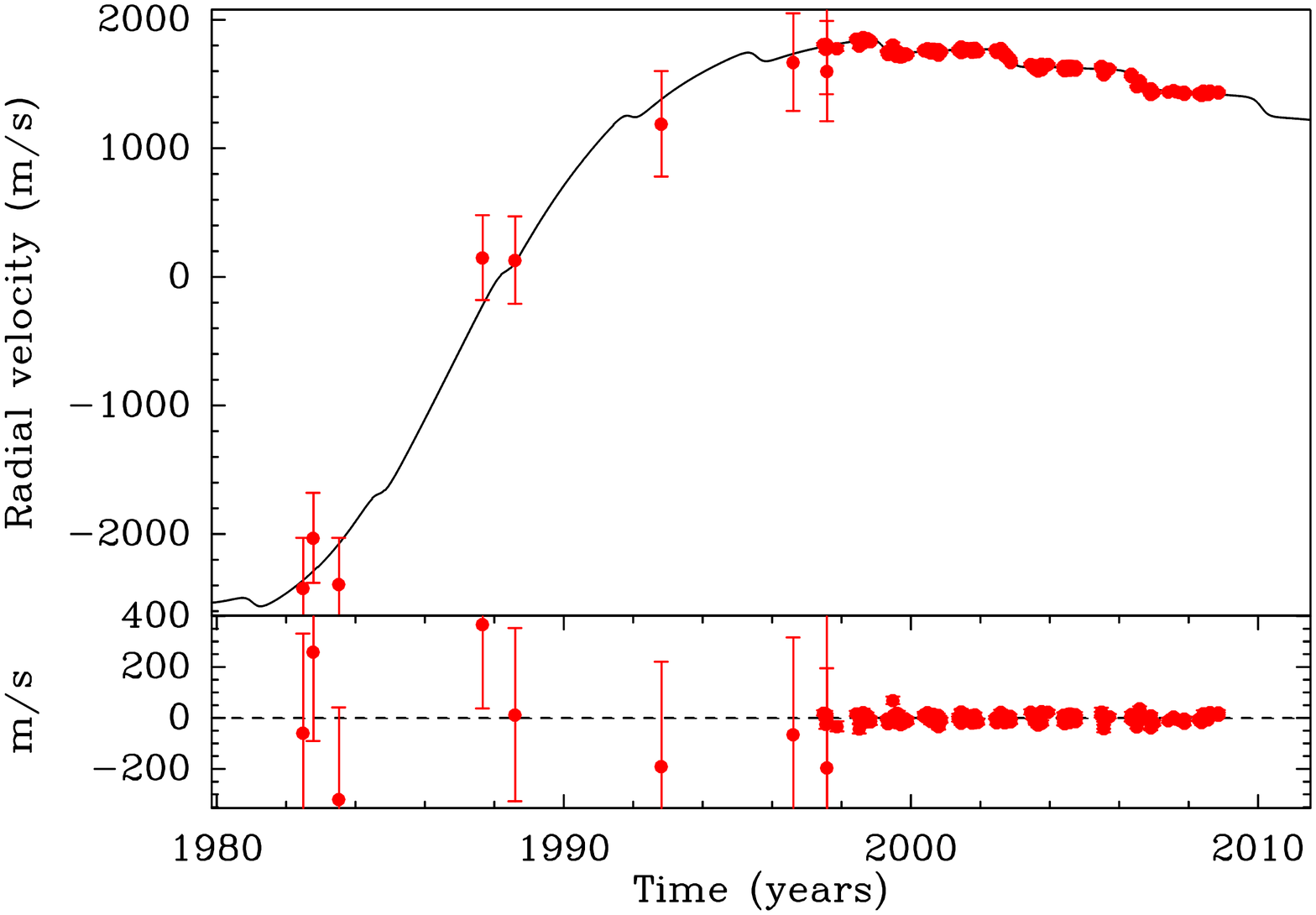}
\includegraphics[width = 8.9cm]{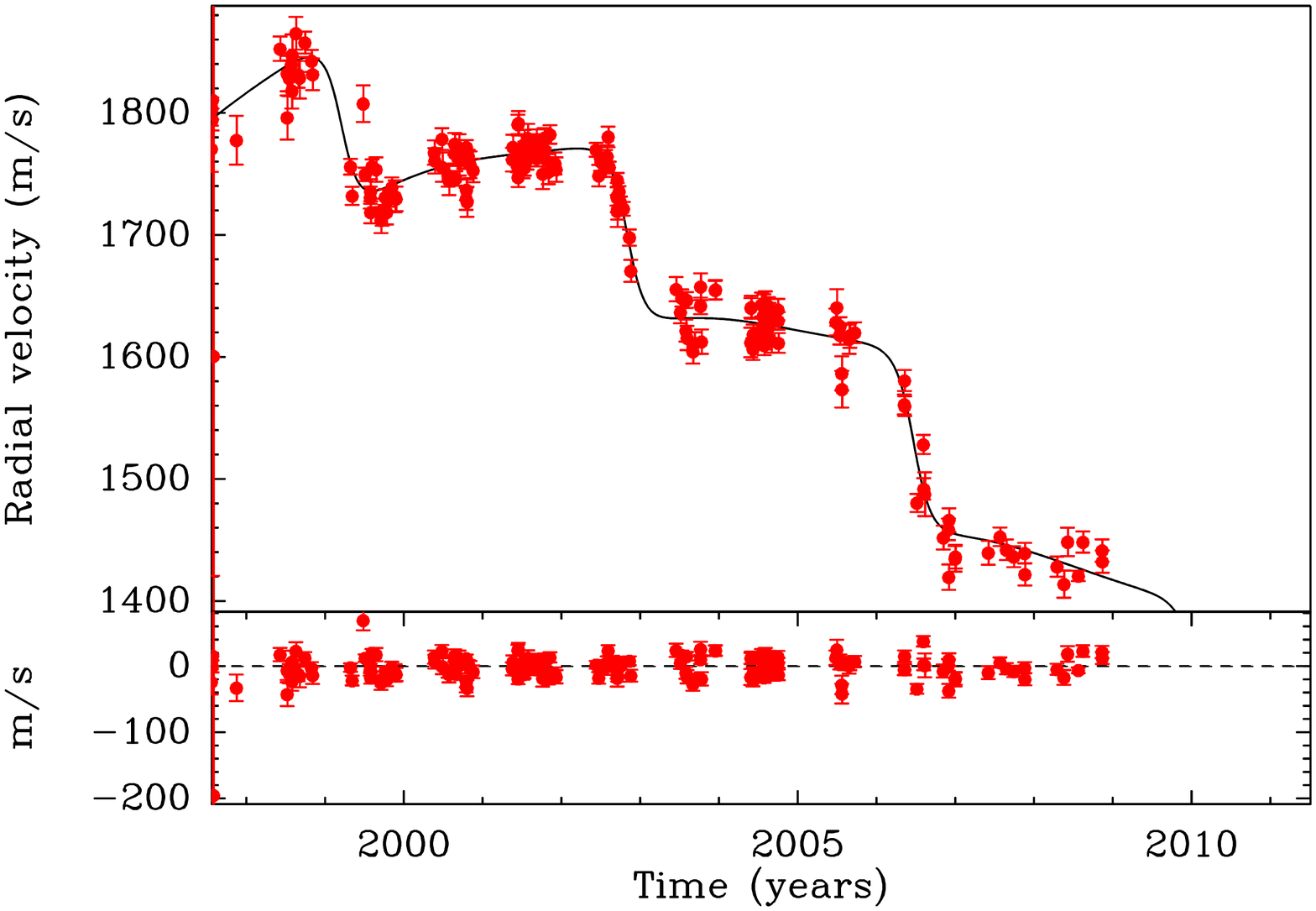}\\
\includegraphics[width = 8.9cm]{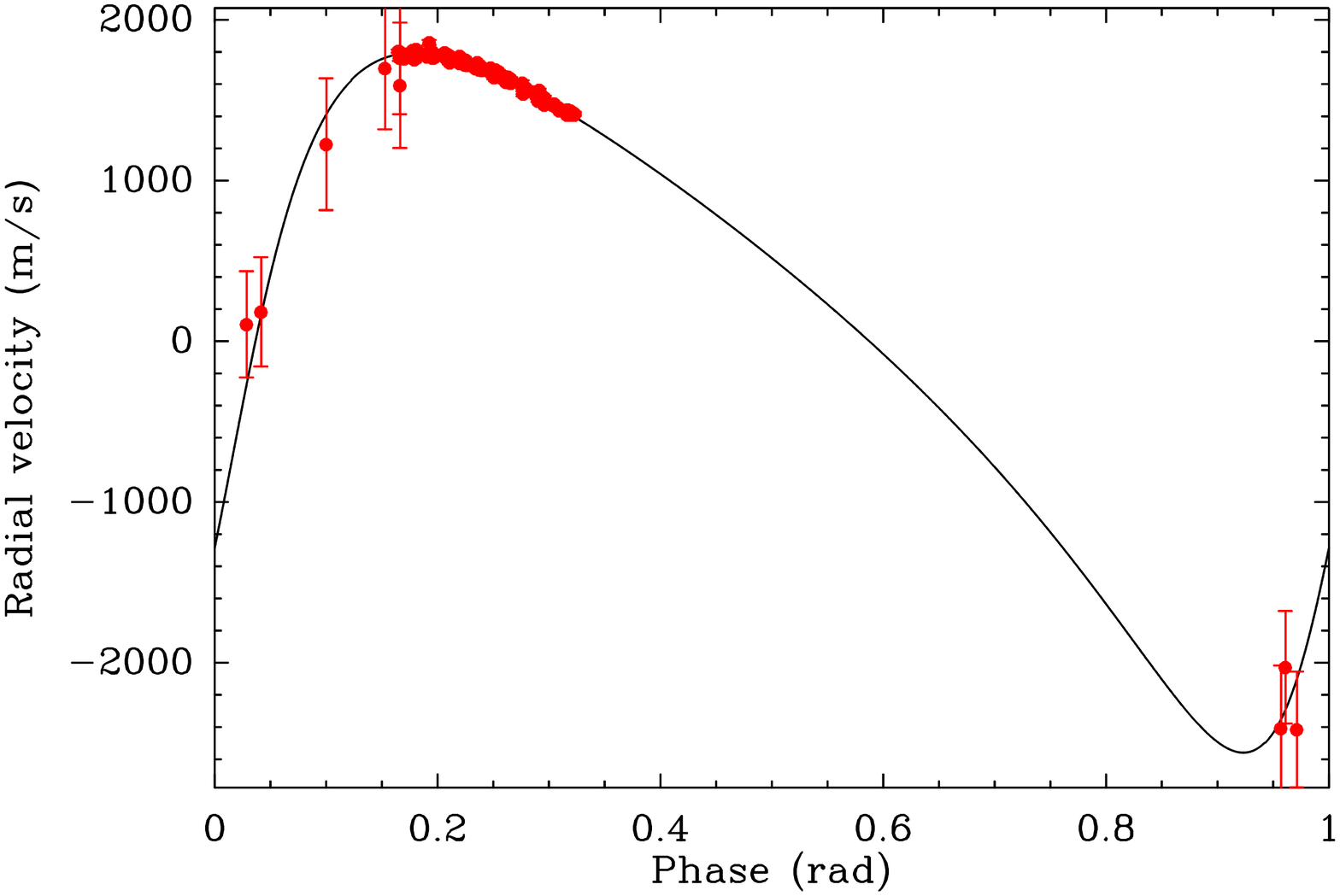}
\includegraphics[width = 8.9cm]{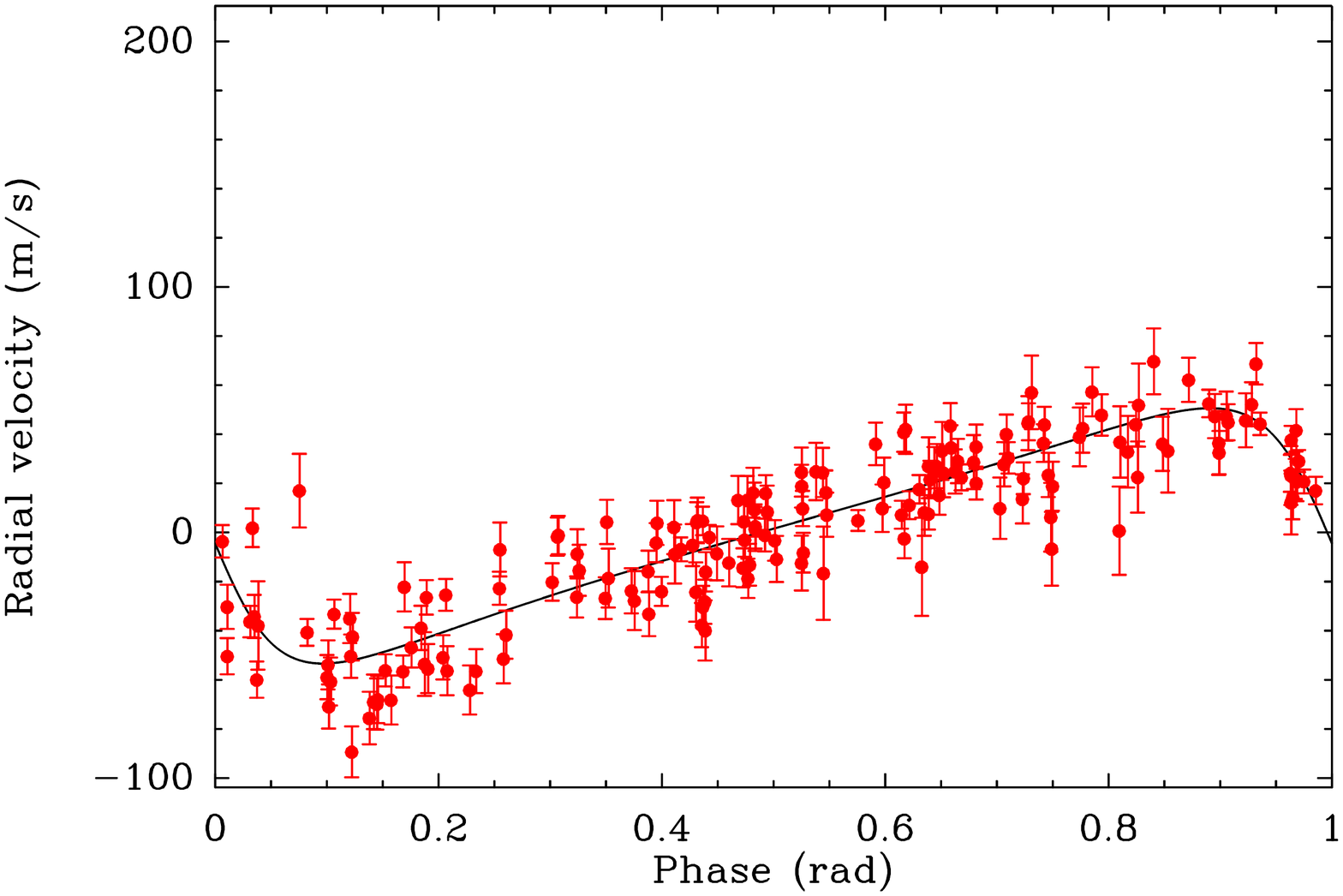}
\caption{Radial velocity observations and orbital solution of
  HD\,196885\,AB based on the CORAVEL, ELODIE, CORALIE and Lick
  surveys. \textbf{Top Left,} Radial velocity variation as a
  function of time with the best orbital solution overplotted
  presenting the complete set of RV observations. \textbf{Top Right,}
  Zoom-in on the RV variation due to the giant planet HD\,196885 Aa
  superimposed to the RV drift due to the companion
  HD\,196885\,B. \textbf{Bottom Left,} Radial velocity variation of
  HD\,196885\,A due to the B component only as a function of the
  orbital phase. \textbf{Bottom Right,} Radial velocity variation
  of HD\,196885\,A due to the Ab planet only as a function of the
  orbital phase. }
\label{fig:orbitfit2}
\end{figure*}

In attempt to constrain the physical and orbital properties of the
HD\,196885 system, we considered simultaneously the astrometric data
points listed in Table~\ref{tab:astro} and all data from past RV
surveys: CORAVEL (9 RV measurements from June 1982 to August 1997),
ELODIE (69 measurements from June 1997 to August 2006), CORALIE (33
measurements from April 1999 to November 2002) and Lick (75
measurements from 1998 to 2008) spanning over 26 years (see Correia et
al. 2008, Fischer et al. 2009).  We used the iterative
Levenberg-Marquardt $\chi^2$ minimization method (Press et al. 1992)
to simultaneously fit both set of data and derive the
characteristics of both orbits (the outer AB orbit between HD\,196885\,B
and A, and the Ab orbit between the giant planet and
HD\,196885\,A). The basic assumption is that both orbits contribute to
the RV signal and that only the AB orbit is seen in the astrometric
motion. For the RV data, separate zero point velocities have been
fitted for each set of data to account for instrumental shifts.  The
fitting routine turned out to rapidly converge towards a unique best
solution reported in Table~\ref{tab:orbit}. This solution is also
sketched in Figs.~\ref{fig:orbitfit1} and \ref{fig:orbitfit2} and
superimposed to the astrometric and RV data points.

With a minimum mass of $M_P\,\rm{sin}i = 2.98$~$M_{\rm{Jup}}$, a
period of $P=3.63$~yr and an eccentricity $of e=0.48$, the solution
for the giant planet (Orbit Ab) is consistent with the orbital
parameters found by Correia et al. (2008) and Fischer et
al. (2009). \textbf{The imaging data lift the degeneracy} of orbital
parameters for the stellar companion HD\,196885\,B. The mass of
$0.45~M_{\odot}$ is in agreement with the companion photometry and the
M1$\pm$1V spectral type (Chauvin et al. 2007). With a period of
$P=72.1$~yrs, our observations confirm a close orbit with a semi-major
axis of just 21~AU. Together with Gl\,86 (Queloz et al. 2000; Lagrange
et al. 2006), $\gamma$ Cep (Hatzes et al. 2003; Neuh\"auser et
al. 2007) and HD\,41004 (Zucker et al. 2004), this system is among the
closest binaries with one component hosting a giant planet.

\begin{table}[t]
\caption{Fit and orbital parameters of HD196885\,Ab and B}
\label{tab:orbit}
\centering
\begin{tabular}{llll}     
\hline\hline\noalign{\smallskip}
 Param.      & [unit]           & HD196885\,Ab       & HD196885\,B      \\
\noalign{\smallskip}\hline\noalign{\smallskip}
$P$          & [yr]             & $3.63\pm0.01$      & $72.06\pm4.59$   \\
$e$          &                  & $0.48\pm0.02$      & $0.42\pm0.03$ \\
$\omega$     & [deg]            & $93.2\pm3.0$       & $-118.1\pm3.1$   \\
$\Omega$     & [deg]            &                    & $79.8\pm0.1$   \\
$i$          & [deg]            &                    & $116.8\pm0.7$  \\
$t_P$        &                  & $2002.85\pm0.02$ & $1985.59\pm0.39$\\
$a$          & [AU]             & $2.6\pm0.1$        & $21.00\pm0.86$     \\  
$M_{Ab}\rm{sin}i$ & $[M_{\rm{Jup}}]$ & $2.98\pm0.05$  & \\     
$M_{B}$      & $[M_{\odot}]$          &                & $0.45\pm0.01$\\   
\noalign{\smallskip}\hline\noalign{\smallskip}
$v_0$ (Coravel) & [km/s]        & \multicolumn{2}{c}{$-31.89\pm0.20$}  \\
$v_0$ (Elodie) & [km/s]         & \multicolumn{2}{c}{$-31.99\pm0.04$}  \\
$v_0$ (Lick) & [km/s]           & \multicolumn{2}{c}{$-1.64\pm0.04$}   \\
$v_0$ (Coralie) & [km/s]        & \multicolumn{2}{c}{$-32.00\pm0.04$}  \\
$\chi^2$        &               & \multicolumn{2}{c}{$465.92$}         \\ 
\noalign{\smallskip}\hline
\end{tabular}
\end{table}

\section{Dynamical evolution and stability}

We performed a preliminary numerical study of this system using the
symplectic N-body package HJS (Beust 2003) \textbf{optimized for}
hierarchical multiple systems. The study was started from the fitted
solution of Table~\ref{tab:orbit}, considering that some free
parameters in the Ab orbit remained (namely the inclination $i$ and
the longitude of node $\Omega$). Depending on the inclination assumed,
the mass of the giant planet could range between the \textbf{minimum}
value quoted in Table~\ref{tab:orbit} and several tens of Jupiter
masses. The angle $\Omega$ had no effect on the RV signal of the
planet, but is related to the mutual inclination $i_r$ (mutual
inclination between the two orbits) by:
\begin{equation}
\cos i_r=\cos i\cos i'+\sin i\sin i'\cos (\Omega - \Omega')\qquad,
\end{equation}
$i'$ and $\Omega'$ are the corresponding parameters for the AB
orbit (listed in Table~\ref{tab:orbit}). $i$ and $i_r$ turn out to be
the more relevant parameters to dynamically characterize the
system. We thus performed various integrations over $10^7\,$yr with
different values of $(i,i_r)$ couples. \textbf{Not all pairs
were compatible with our constraints.}  

\begin{figure*}[t]
\makebox[\textwidth]{
\includegraphics[width=0.48\textwidth]{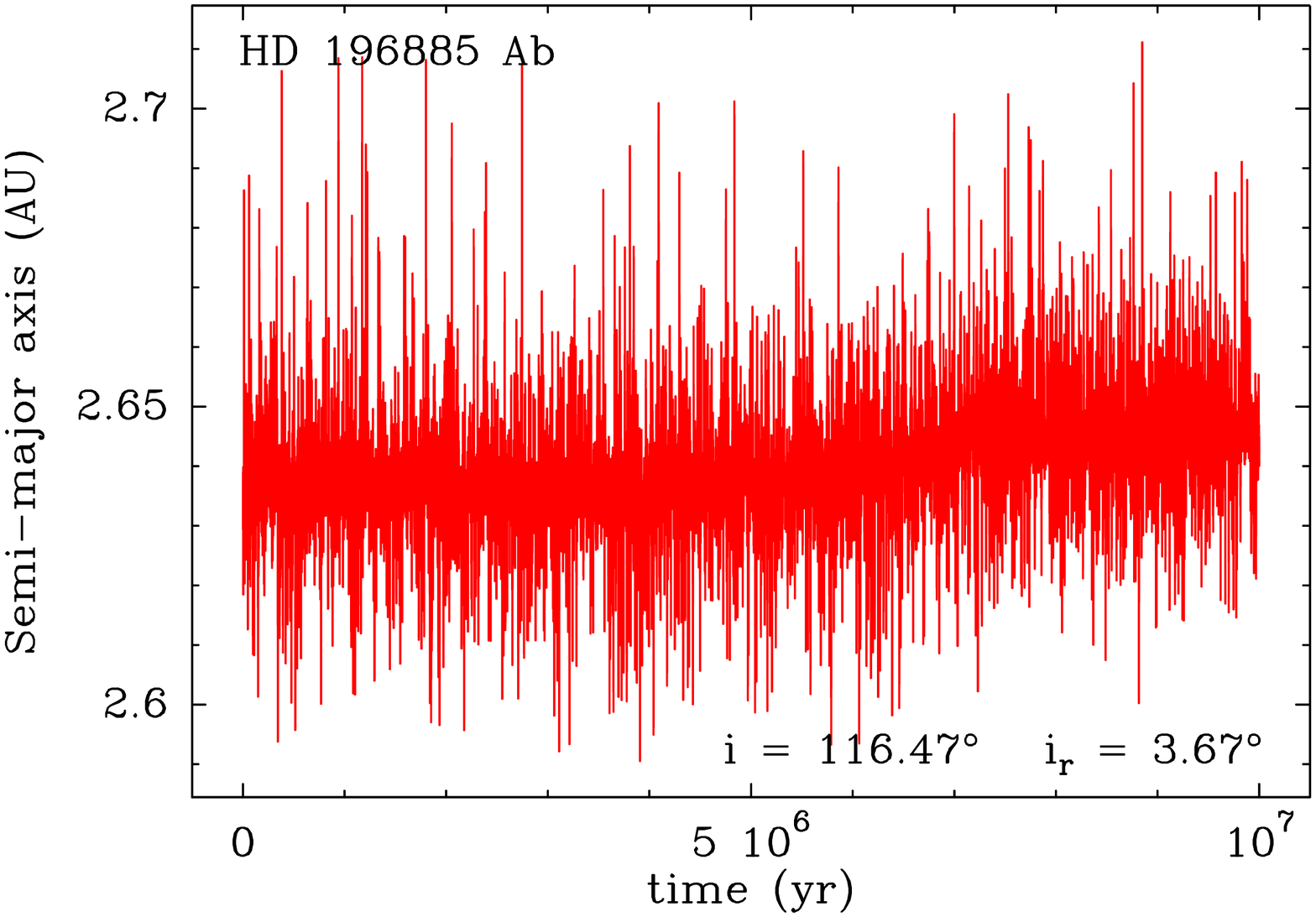}\hfil
\includegraphics[width=0.48\textwidth]{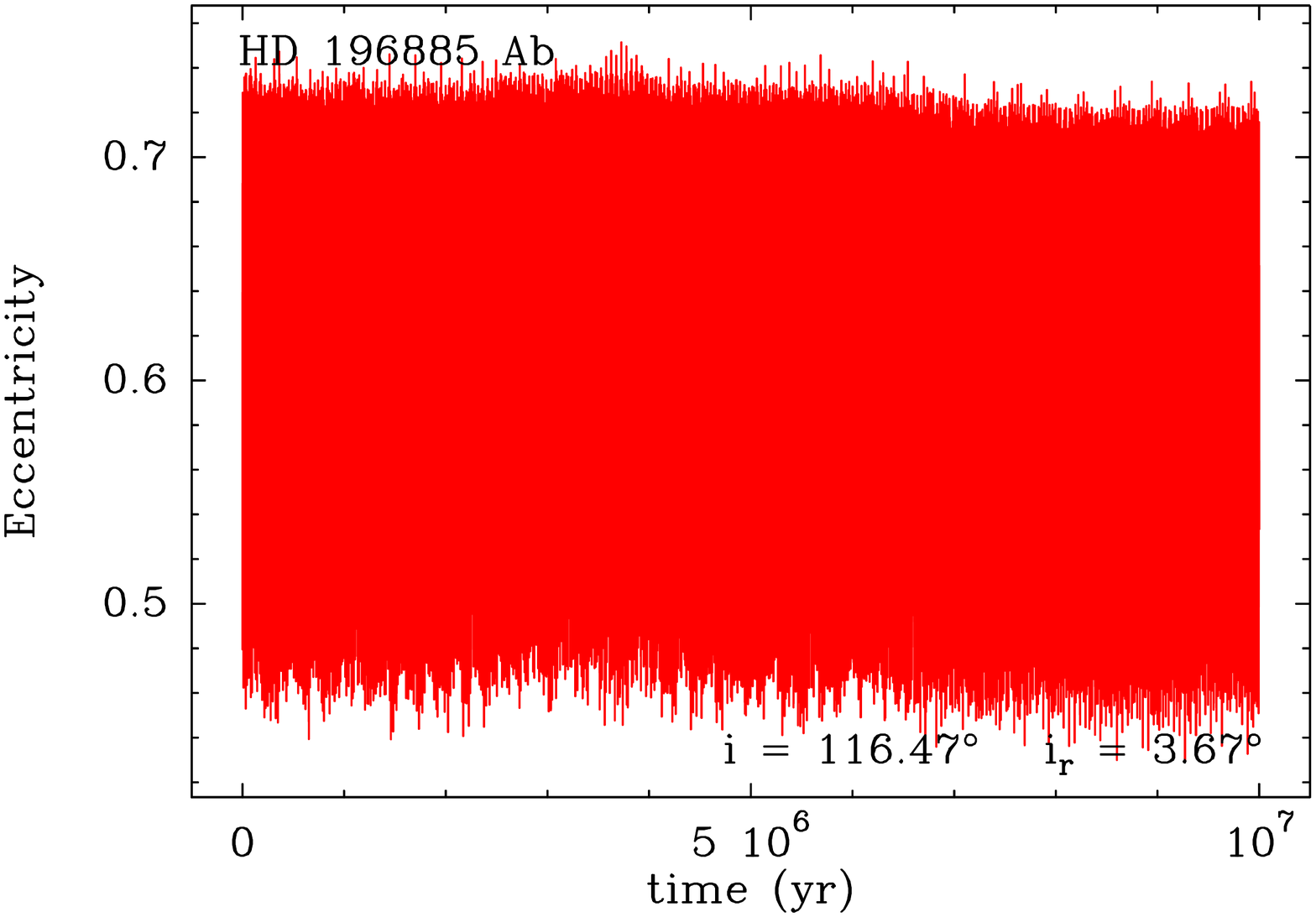}}
\makebox[\textwidth]{
\includegraphics[width=0.48\textwidth]{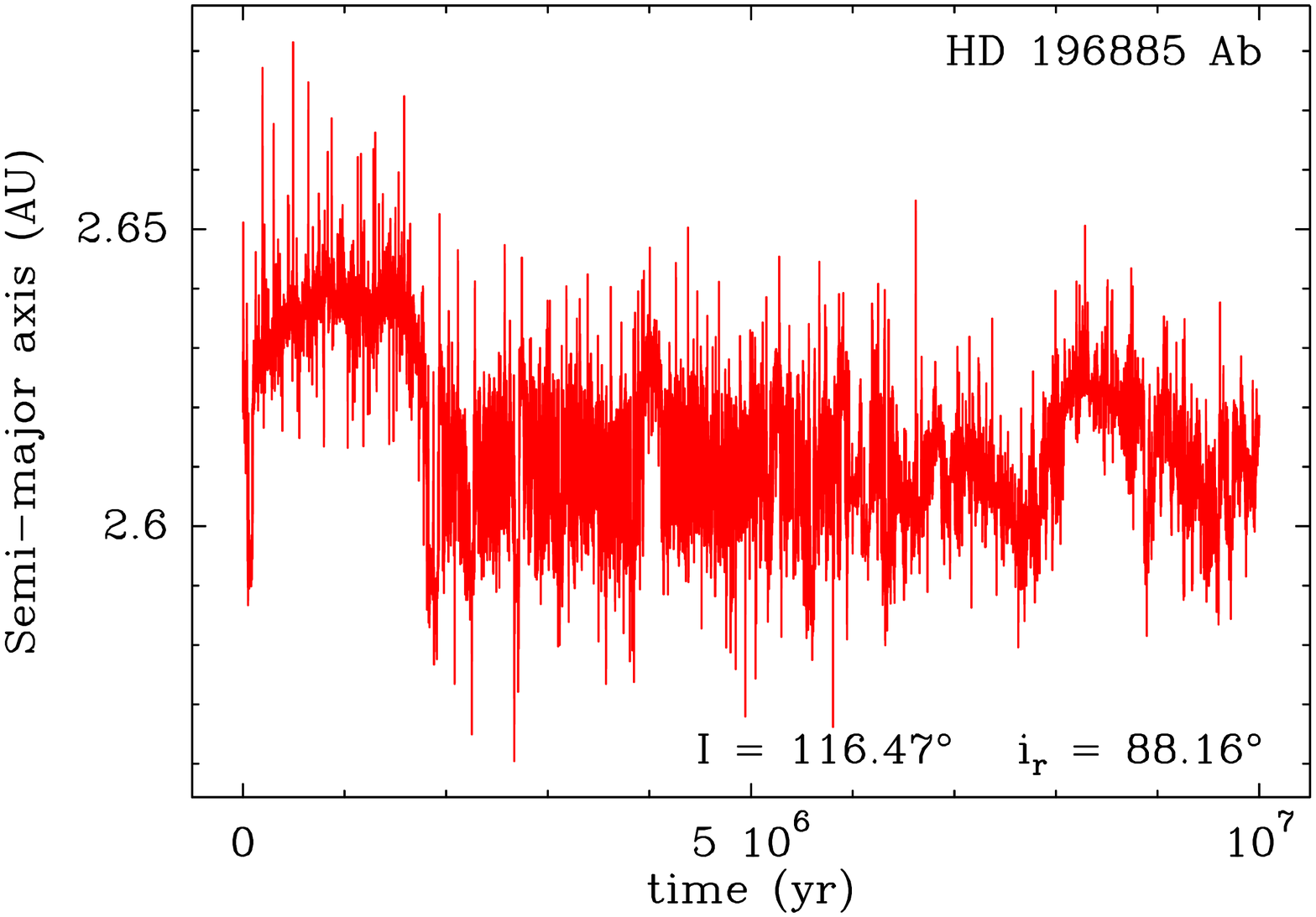}\hfil
\includegraphics[width=0.48\textwidth]{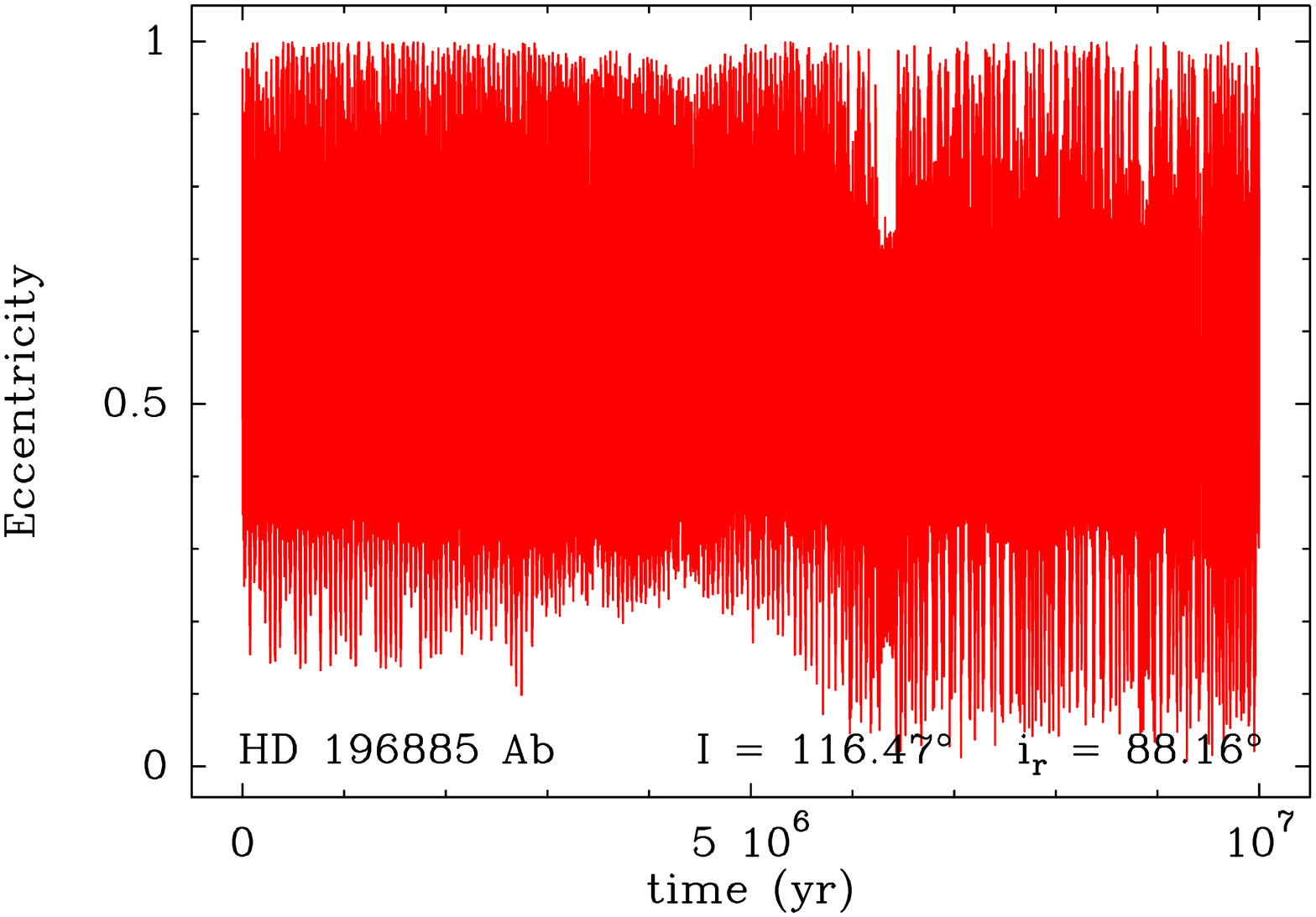}}

\caption{Examples of integration runs for the HD 196885 3 body
  system. The evolution of the inner orbit Ab is shown over
  $10^7\,$yr. \textbf{Top,} a typical coplanar solution, with semi-major
  axis (left) and eccentricity (right) evolution; \textbf{Bottom,} same
  for a typical highly $i_r$ solution with kozai oscillations (see
  eccentricity).} 

\label{fig:orbevol}
\end{figure*}

Two main results were obtained from our simulations. The first result
shows that in all cases the system appears chaotic with significant
(more or less erratic) changes of semi-major axis for the Ab
orbit. The stability of the system needs to be therefore investigated
over a much longer timescale. The second result shows that the mass of
the giant planet HD\,196885 Ab (hence the Ab orbit inclination $i$)
does not significantly affect the stability of the system. Additional
simulations with similar $i_r$ but different $i$ show similar
behaviours. This can be explained by the dominant influence of the
higher mass companion HD\,196885\,B as the giant planet behaves almost
like a massless test particle in this system. The mutual inclination
$i_r$ is however a key parameter for the system stability. The
behaviour of two different solutions (low and high $i_r$) concerning
the Ab orbit is shown on Fig.~\ref{fig:orbevol}. The AB orbit remains
slightly affected and is not considered here. In both cases, we
note significant semi-major axis oscillations revealing a chaotic
regime. The eccentricity oscillations are high, especially in the high
$i_r$ case. In fact, all high $i_r$ configurations fall in the Kozai
resonance regime (Kozai 1962). Under the effect of secular
perturbations by the outer body, the giant planet is subject to a
periodic evolution that drives it to a lower inclination but a very
high eccentricity. This mechanism is active in non-coplanar
hierarchical triple systems (Harrington 1968; Ford et al. 2000) and is
therefore not surprising in our case. It was actually invoked to
explain the high eccentricity of some extrasolar planets in binary
systems (Mazeh 1997; Holman et al. 1997; Libert \& Tsiganis 2009;
Fabricky et al. 2007).

Surprisingly, our results show that the system is more stable in the
Kozai regime. The Kozai resonance is an angular momentum exchange
process. In a pure 3-body system, it does not affect the semi-major
axis evolution and the system stability. Therefore the real level of
instability can be read from the evolution of the semi-major axis,
which is more stable in the Kozai configuration (see
Fig.~\ref{fig:orbevol}). This is a general trend present in all our
simulations. For coplanar systems, only marginally stable solutions
are found.  Consequently, our results suggest that a non-coplanar
configuration, characterized by a Kozai regime, is more probable for
HD\,196885. The high eccentricity of the Ab orbit fit is another
strong indication in favour of a Kozai regime. If confirmed, this
system would constitute one of the most compact non-coplanar systems
known so far. Deeper investigation will be needed to further constrain
this issue and will be presented in a forthcoming paper.

\section{Discussion}

Stellar companions do influence the occurrence and the properties of
planets in circumstellar orbit (i.e orbiting one of the binary
compenent). Observations show that planets seem less frequent in close
($\le100$~AU) binaries (Eggenberger et al. 2007, 2008) and that
massive short-period planets appear to be preferentially found
orbiting one component of a multiple system (Zucker \& Mazeh 2002;
Eggenberger et al. 2004; Desidera \& Barbieri 2007). Among the few
tens of exoplanetary hosts, Gl\,86, $\gamma$ Cep, HD\,41004 and
HD\,196885 are of particular interest as they are short period
binaries with semi-major axis lower than $\sim25$~AU (see system
properties in Table~3). In such close systems, we can wonder how their
planets might have formed and survived to the close
interaction with the outer binary companion. The presence of a binary
companion is expected to truncate the protoplanetary disk, affect its
eccentricity and impact the velocity dispersion and evolution of
planetesimals which could lead to a planet formation hostile
environment (e.g. Th\'ebault et al. 2006; Paardekooper et al. 2008;
Xie et al. 2008; Marzari et al. 2009; Cieza et al. 2009). In the
context of $\gamma$ Cep, Th\'ebault et al. (2004) and Paardekooper et
al. (2008) showed that core-accretion formation of a giant planet was
feasible, although it was probably slowed down (assuming coplanarity
of the disk, planet and binary). Jang-Condell et al. (2008) found that
disk instability was in addition more unlikely as a massive
circumstellar disk and/or extremely high accretion rate as seen in FU
Orionis events are required. Kley \& Nelson (2008) then examined the
dynamical evolution when a protoplanetary core has formed and begins
the accretion phase. They constrained to ($a_p\le 2.7$~AU) the initial
semi-major axis of the giant planet that could lead to a stable
solution surviving inward migration and eccentricity
variation. HD\,196885 closely resembles the $\gamma$ Cep
system. However, the higher mass ratio between the binary companion
and the primary and the further location of the inner planet makes it
even more challenging for core-accretion planet formation theories.

\begin{table*}[t]
\caption{Orbital parameters of \textbf{small separation} ($a\sim20$~AU) binaries with one exoplanet around one component.}
\label{tab:exobinaries}
\centering
\begin{tabular}{lllllllll}     
\hline\hline\noalign{\smallskip}
Name      & Age        & $M_1$    &   $M_B$    &   $a_B$     &    $e_B$    &    $M_p$     &    $a_p$     &   $e_P$          \\
          & (Gyr)           &  ($M_\odot$)&  ($M_\odot$) & (AU)        &             &  ($M_{Jup}$)     & (AU)         &                  \\
\noalign{\smallskip}\hline\noalign{\smallskip}
HD\,196885   &  1.5-3.5 & 1.3     & 0.45       &   21.0          & 0.42        &     3.0       &   2.6       &   0.48     \\ 
$\gamma$ Cep &  6.6     & 1.6    & 0.41       &   19.0          & 0.41        &     1.6       &   2.0       &   0.12     \\ 
Gl\,86       &  2.0     & 0.8     & 0.45       &   $18.0^a$       & 0.40        &     4.0       &   0.11      &   0.05           \\ 
HD\,41004    &  1.6     & 0.7     & $0.40+0.02^b$       &   $\sim20^b$       & 0.40        &     2.5       &   1.64      &   0.39           \\ 
\noalign{\smallskip}\hline
\end{tabular}
\begin{list}{}{}
\item[$^{\mathrm{a}}$] predicted intial semi-major axis of 13~AU for the binary (Lagrange et al. 2006).
\item[$^{\mathrm{b}}$] HD\,41004\,B is actually a SB1 binary composed of a M4V star and a $M_{Bb}=19$~$M_{Jup}$ brown dwarf (Santos et al. 2002; Zucker et al. 2004).
\end{list}
\end{table*}

Although coplanarity seems a reasonable assumption for the planet
formation in close binaries (Hale 1994), our preliminary dynamical
study shows that a non-coplanar configuration, characterized by a
Kozai regime, is more probable for HD\,196885. Its origin might
therefore be questioned, as it seems difficult to build a non-coplanar
system from a single disk. The giant planet has to form and survive in
a non-coplanar system under the combination of the secular pertubation
of the close binary companion and disk evolution. This
has to be tested. An alternative scenario could be that the binary
was less compact in the past. Close encounters with other young stars
could have led its orbit to shrink (Malmberg et al. 2008). The
companion may also have been captured after the formation of giant
planet around the primary in the birth cluster (Portegies Zwart \&
McMillan 2005; Pfahl \& Muterspaugh 2006). We could even think of a
capture of the giant planet itself.  These are of course speculations
but the high present-day eccentricity of HD\,196885\,B is also an
indication of a chaotic past dynamical history of that system. Further
investigation will be needed to further constrain and test the system
stability and origin on a broader parameters range.

\section{Conclusion}

We have reported the results of four years of astrometric monitoring
of the planet-host binary system HD\,196885\,AB using NaCo at
VLT. Combined with RV observations, our imaging results enabled to
derive the inclination, the eccentricity and the true mass of the B
component, a low mass star orbiting at 21~AU. We found consistent
solutions for the planet with results previously reported in the
literature. We also confirmed that HD\,196885\,AB belongs to the rare
cases of close binaries with one component hosting a giant planet,
offering an ideal labolatory to study the formation and evolution
processes in such extreme planetary systems. Finally, we have run
N-body numerical simulations to test the system stability. The main
result suggests that non-coplanarity and high mutual inclination,
characterized by a Kozai regime, favor the system stability. If
confirmed, HD\,196885 would constitute one of the most compact
non-coplanar system hosting a circumstellar planetary system. How
planet formation could have occured in such an extreme and hostile
environment remains a challenging question. In-situ formation of a
non-coplanar system cannot be excluded and has to be tested. However,
alternative scenarii such as an external perturbation with stellar
encounters modifying the binary properties or capture mechanisms are
likely to offer reasonable explanations.

\bibliographystyle{aa}

\begin{acknowledgements}

We would like to thank the staff of ESO-VLT for their support at the
telescope, the Osservatorio di Arcetri and \textbf{its} director Francesco
Palla for the visiting scientist stay of the author during the writing
of this paper.  We would like also to thank Philippe Th\'ebault for
his comments and suggestions to the article. Finally, this publication
has made use of the SIMBAD and VizieR database operated at CDS,
Strasbourg, France. And, we acknowledge partial financial support from
the {\sl Programmes Nationaux de Plan\'etologie et de Physique
  Stellaire} (PNP \& PNPS) and the {\sl Agence Nationale de la
  Recherche}, in France. AE is also supported by a fellowship for
advanced researchers from the Swiss National Science Foundation (grant
PA00P2\_126150/1).

\end{acknowledgements}

\end{document}